# A Novel Strategy Selection Method for Multi-Objective Clustering Algorithms Using Game Theory

Mahsa Badami, Ali Hamzeh, Sattar Hashemi
School of Electrical and Computer Engineering, Shiraz University, Shiraz, Iran

**Abstract**
The most important factors which contribute to the efficiency of game-theoretical algorithms are time and game complexity. In this study, we have offered an elegant method to deal with high complexity of game theoretic multi-objective clustering methods in large-sized data sets. Here, we have developed a method which selects a subset of strategies from strategies profile for each player. In this case, the size of payoff matrices reduces significantly which has a remarkable impact on time complexity. Therefore, practical problems with more data are tractable with less computational complexity. Although strategies set may grow with increasing the number of data points, the presented model of strategy selection reduces the strategy space, considerably, where clusters are subdivided into several sub-clusters in each local game. The remarkable results demonstrate the efficiency of the presented approach in reducing computational complexity of the problem of concern.

## 1. Introduction

Nowadays, multi-objective clustering is a well-stablished field growing rapidly in many domains. In recent years, several new practical applications need object clustering at various levels with multiple criteria [1]. Cross-disciplinary application areas such as emergency resource deployment, ad-hoc networks, and facility location, require often several conflicting metrics optimization, such as compaction and equi-partitioning [2]. Therefore, there is an indispensability to develop a promising technique for simultaneous optimization of conflicting objectives, while a single-objective clustering methods, such as K-means, focus on compaction and identify clusters which may not be equi-partitioned. Hence, a multi-objective method, that provides these objectives, suggests much better clusters. The final clusters are compact while they support almost equal data points.

Recently, a novel approach is developed in order to solve the latter problem by Gupta and Ranganathan [2]. This algorithm comprises three components: 1) initial step which includes an iterative hill-climbing-based partitioning, 2) a multistep normal form game formulation that identifies the initial clusters as players and resources on the basis of certain properties, and 3) a Nash Equilibrium to evaluate optimal clusters. The presented method by Gupta and Ranganathan, so-called GTKMeans, achieves significant results. However, GTKMeans suffers high time and game complexity of large data sets due to growth of number of players, strategies and consequently payoff matrices. In order to solve this problem, Gupta and Ranganathan proposed an ensemble-based method (PKGame) in order to reduce time complexity via elimination of iterations. However, it does not effect on game complexity of inside of each local game. In order to deal with the presented challenge, we present a novel approach to reduce strategies set for each player. In this case, the size of payoff matrices is reduced effectively, resulting to improve the algorithm to be applied on large data sets.

In the next section, we briefly review existing clustering techniques and various application domains of game theory. Section III, explains strategy definition which specifically describes the proposed strategy selection method. Next in Section IV, the experimental results for the performance of the algorithm on both real and artificial data sets are presented. Moreover, the proposed approach is analyzed on basis of two fairness metrics as well as time and game complexity. Eventually, the conclusion is discussed in Section V.

## 2. Related Works

So far, numerous Multi-Objective clustering algorithms have been reported in the literatures. Traditionally, Multi-objective methods have been categorized as ensemble, evolutionary and microeconomic methods. Cluster ensemble frameworks combine different partitions of data using consensus functions [3]. Clustering ensembles have been provided as a powerful method rather than individual clustering methods. However it is not able to support simultaneous objective optimization [4]. In other hand, evolutionary computing algorithms such as MOEA [5], PESA-II [4] and MOCK [5] provide solutions which are strictly better than other solutions, known as Pareto front. These methods identify better clusters than ensemble clustering, where it optimizes objectives concurrently [6].

New generation of multi-objective methods is microeconomic methods which raise many interest in recent years [2][7][8][9][10][11]. Classical game theory deals with rational individuals, players, who play game with each other. Each player defines a set of options or strategies, and then, he plays based on his strategies, in order to maximize his utility or payoff. The payoff scores are defined based on strategies of other players who in turn try to maximize their own payoff. The game is ended over by finding an equilibrium that all players create their beliefs based on an estimation of what others might do (strategic thinking). Generally, equilibrium makes a choice based on the best response of any player. The key idea of a game theoretic framework is the social fairness which ensures that every player is satisfied with respect to every other player in the system [9].

Microeconomic techniques have been widely researched in different domains of computer science as diverse as computer vision and bioinformatics. Moreover, the game-theoretic clustering framework falls into the class of similarity-based approach [9]. While a cluster is a set of mutually similar objects, a clustering problem can be solved via game theory concepts. Several clustering methods are proposed based on game theory, namely, a pair wise clustering for overlapping groups [12] ,[8], hyper-graph clustering [9]. One of the latest methods, which clusters spatial data based on the concepts of microeconomic theory, so called GTKmeans, was proposed in [2]. This approach models the clustering problem as a normal form of non-cooperative game in order to optimize both compaction and equi-partitioning objectives, simultaneously. Compaction is implemented on the basis of minimizing Sum Squared Error (*SSE*) with sum square Euclidean distance. Alternatively, equi-partitioning is implemented by minimizing sum of squared load values (L) formulated as follow:

$$L = \sum_{k=1}^{K}(l_k - l_{ideal})^2 \quad (1)$$

GTKMeans provides three important steps: First step of algorithm is started with a single iteration of K-means which serves initialization clusters [2]. If a single iteration of k-means results in equi-partitioned clusters, centers are updated, and they are served for the next iteration of k-means. In the second step, if the cluster centers were not equi-partitioned, a new game is needed to be formulated in order to make their size equal. The game defines clusters with more units than ideal size of units as resources and identifies those clusters that contained less data units than ideal size of units as players. The ideal size of a cluster is defined as $L_{ideal}=N/K$, where *N* is total number of data, and *K* is the predefined number of clusters. Afterward, the strategies set are established for each player based on the required units from resources. Each strategies set is developed as the number of returned units to resource in order to keep its units beyond ideal size or at least equal to it. The game is continued by payoff calculation that models the gain or loss of the player with respect to other players' strategies. The payoff matrix is formulized based on both compaction and equi-partitioning objectives. The payoff function is modeled as a geometric mean of the total loss incurred by a player in terms of the difference between the SSE before and after the rival players plays their strategies, and the absolute value of the equi-partitioning metric, corresponding to the strategy of the current player. At the third step, a Nash Equilibrium solution is identified from the payoff matrix in order to determine the final strategies set. According to this strategies set which consist of one strategy for each player, a temporary reallocation of data objects is performed. If the reallocations improve the overall objectives, they are made permanent, and the cluster centers are updated. The new clusters are served for another iteration of K-means. The procedure is repeated until the stopping criterion is satisfied.

One Particularly important issue is that Strategies set for each player is arise by growth of the number of players during each local game, and number of each game is greatly dependant on number of all players and resources. In fact, it increases with number of clusters as well as number of data objects. Therefore, as GTKMeans, the proposed methodology is ideally suited for multi-objective clustering in small to medium sized data sets. As shown in [2], an ensemble-based algorithm, named PKGame, is required, due to high complexity of large data sets. The ensemble-based method is performed in two steps. At the first step, a full iterative K-means is performed, and at the second one, a game is modeled. However, the point is, the game is established only once for each conflicting resource center with its players. Although it reduces over all time and game complexity, it does not effect on complexity of each local game. In order to solve this problem, we present a new method to select some strategies instead of all of them for each player in each local game. In this case, size of payoff matrices are decreased significantly which cause less time complexity, consequently. Therefore, practical problems, with more data points, become intractable with less computational complexity. The following section describes details of presented algorithm.

## 3. The Proposed Method

The most significant aspect of game theory is definition of strategies used in each game, since it determines the player's fate in the game. Since the notation of strategy is an important factor in determining the computational complexity of the model, it affects payoff matrix size directly. Therefore, eliminating some strategies from each player's strategy set can significantly improve computational complexity of the model.
In this section, first, a detailed description of the formulation the strategies is explained. Next, we describe our novel idea about strategy selection and present a new method to choose a subset of input strategies for each player.

### 3.1 Strategy Profile Generator

Basically, the strategy set proceeds in two steps: at first step, each player finds his closest resource based on the Euclidean distance of the cluster centers, and then requests necessary units from a closer resource. Sometime, a particular situation may occur in consequence of this process. The resource may allocate more units to a player or players than its available overhead units, the units more than $l_{ideal}$. In this case, the resource does not have enough overhead units to allocate. Hence, a new game is formulated to solve this situation, in order to optimize overall objectives. Step two is held in each local game. During this step, each player can release given units, in order to ensure that conflicted resource is consistent, in other words, it is in equi-partitioned state.

For more description, consider a local game with two players, $p_1$ and $p_2$, and one recourse, $r_1$. , $p_1$, $p_2$ and $r_1$ have 4, 1 and 8 units, respectively. $p_1$ and $p_2$ send their request to their closest resource, $r_1$, while $l_{ideal}$ is 7, strategies sets for $p_1$ and $p_2$ are {0, 1, 2} and {0, 1, 2, 3, 4, 5}, respectively.

3.2 Strategy Selection Method

In this section, we present our new approach in order to select a subset of strategies for all players. In this approach some strategies are selected.

As explained in previous part, strategies set for each player consists a set of number. These numbers illustrate number of units which should be returned to resource to make it equi-partitioned. Here, we apply a kind of sub-clustering method in each local game over each conflicted resource. Therefore, we replace each unit with a group of units. In other words, units transferred in groups rather than individuals.

If a conflicted resource has $N_r$ units, number of sub-clusters is $N_r / N_s$, which each of them has $N_s$ units. Based on it, when a player wants to receive units, he gets one cluster which has $N_s$ closest unit to the player. We implement this idea via strategy definition.

To clarify it better, consider the local game from previous example. Player $p_i$ whose strategies set is {0, 1, 2, 3, 4, 5} changes his to {0, 2, 4, 5} since $N_s$ is 2. It means that we consider only numbers which are multiple of $N_s$. In other words, we create sub-clusters which each has two units and if there is a need to transfer a group, two closest units are conveyed. In the case that strategies end with an odd number, strategy selection considers the last one too.

$N_s$ is based on size of strategies sets in each local game. We can define $N_s$ adaptively; however, we consider predefined value for it with the aim of simplicity.

Although strategies set may enlarge with growth equipartitioning metric ($L$), this model of strategy selection reduces the strategy space, considerably, where clusters is subdivided into several sub-clusters in each local games.

## 4. The Experimental Result

In this section, to explore the ability of the proposed method a series of experiments were conducted of real-world and artificial data sets. The performance of the proposed algorithm also has been evaluated in terms of computational complexity as well as objective efficiency.

4.1 Description of Data Sets

We use two kinds of data sets inspired from other studies to experimentally study performance of the discussed method and evaluate the effectiveness. The real-world data sets include British Town Data (BTD) [13] and German Town Data (GTD) [14]. The former consists of 50 British towns' descriptions that associated with four principal socioeconomic features. Alternatively, the latter one is a two-dimensional data set contains location coordinates of 59 German towns. The second types of datasets are developed for better evaluation of the proposed strategy selection. It is generated with a random set of data on two dimensions, so called DS1. These random points are normally distributed within dimension space. Their means and variances are varied from $0 \leq \mu \leq 10$ and $\delta = \pm 2$, respectively. 150 data points are generated, and they are going to be partitioned into 4-8 clusters.

All experiments are run on 2.00 GHz Intel core 2 Duo CPU with 2.50 GB RAM.

4.2 Results and Discussion

We study the performance of the proposed strategy selection method in terms of computational complexity. This complexity is defined by different features: number of players, number of strategies, size of the payoff, type of the game, type of the equilibrium and many other factors. Here, we consider the complexity in both time and game complexity. The latter one determines the size of the game based on both strategies sets and payoff matrices size. Moreover, we show that how both compaction (*SSE*) and equi-partitioning measure (*L*) act while we apply strategy selection on GTKMeans, PKGame.

We perform GTKMeans and PkGame with and without strategy selection method, over GTD on 50 iterations. All the algorithms have same random initialization points; thus the comparison is quit fair.

Fig. 1 displays average strategies sets of the algorithms per number of players on different cluster size (K=4, 5, 6,7 and 8) while Fig. 2 demonstrates payoff matrices size of them. These figures depict the evaluation of GTKmeans and PKGame while they are using strategy selection with different value for $N_s = 2, 3$. As the numbers of clusters increase, the potential number of players, and consequently the strategies increase; then the game gets large. However, strategy selection offers a remarkable reduction over game complexity while it does not hurt *L* and *SSE*, as it can be seen from Fig. 3 These comparative graphs demonstrate average improvement in *SSE* and *L* over initial clusters for the algorithms before and after applying strategy selection method. In order to evaluate time complexity, Table 1 shows average execution time, in seconds, for different number of clusters. As it is observed in Table 1, the executive time in both game-theoretic algorithms decreases significantly while they use the strategy selection method. It can be infers that the elegant strategy selection method, reduce number of strategies which cause reduction on payoff size and overall game execution time. Based on the presented result, GTKMeans using strategy selection has less computational complexity and more performance in comparison with PKGame.

Similarly, the average performance of new methodology also is examined on the BTD data set. In the experiment, we assess the proposed strategy selection with $N_s = 2, 3$ on PKGame. Fig. 6 and Fig. 7 illustrated the superiority of our elegant method on the basis of both time and game complexity. These figures compared with Fig. 4 prove that the strategy selection method reduces computational complexity of PKGame, significantly, while it has high efficiency on *L* and *SSE* improvements. In addition, Table 2 shows average time complexity of PKGame using the strategy selection method.

Due to better comprehension, the simulations are performed on DS1 with different number of points. In this case, we are able to illustrate the effect of strategy selection over game-theoretic multi objective methods while medium and large sized data sets are used. Although, Strategy selection is examined with $N_s =2, 3$ and $4$, it can be tunable based on assumption over size of payoff matrices. Fig. 8, Fig. 10 and Table 3 proves that the strategy selection with high $N_s$ has less complexity as well as high performance in both SSE and L, shown in Fig. 5.

Based on the presented results both the game complexity and the time complexity increase in the number of players and strategies once the number of data objects and number of clusters grow. However, we can control this effect with applying the strategy selection method. It is worth to notice that strategy selection decrease time and game complexity without reduction on L and SSE improvement, as it is observed from given results.

4.3 Efficiency and Scalability

In the previous section, we demonstrated the supremacy of the presented strategy selection on the basis of both game and time complexity. In this section, we are going to show how this advancement effect on fairness. A fairness measure is used to determine whether users or applications are receiving a fair share of system resources. Jain's Fairness Index [15] and geometric mean index are two appropriate criteria which are suitable. Jain's Index rates the fairness of a set of values; each value is corresponding to improvement of one single objective. The results range from 0 (worst case) to 1(best case).

$$\Im(x_1, x_2, ..., x_n) = \frac{(\sum_{i=1}^{n} x_i)^2}{n \times \sum_{i=1}^{n} x_i^2} \quad (2)$$

In order to more comprehensive analysis, geometric mean index is used to identify the relative improvements in optimization values of the various clustering. The ability of geometric mean index is to obtain criteria as a single index. Geometric mean of $n$ nonnegative numerical values is the $n^{th}$ root of the product of the n values.

$$G = \sqrt[n]{x_1 x_2 \cdots x_n} \quad (3)$$

Each of these values is matched with the improvement of one of our. In this case, best result is 100% when both *L* and *SSE* get 100% improvement.

Both indexes are examined on DS1 for different number of cluster size in order to compare GTKMeans and PKGame with and without using strategy selection. As illustrated in Table 4, results signifies that our strategy selection approach does results better computational complexity, with high Jain's and geometry mean indexes in all clusters size.

## 4. Conclusion and Future Work

In this paper, we have presented a novel strategy selection for multi-objective clustering algorithms, GTKMeans and PKGame, on the basis of the game-theoretic framework. This method optimizes two important metrics, compaction and equi-partitioning. GTKMeans consists of multiple game iterations where a multi-step game is performed in each iteration, while PKGame performs only one iteration of the multi-step game in order to reduce complexity. PKGame is fast since only one set of games is played for each conflicted resource, and there are no other iterations via K-means. However, it has no effect on game complexity. By selecting a subset of strategies, strategies set and payoff matrices size reduce effectively, hence, it causes remarkable reduction over time complexity. This selection is based on a sub-clustering method over each conflicted resource. It is worthy to note that using strategy selection, GTKMeans offers fairer and better clusters with higher performance at the expense of more steps for local games and less time for each of them, compared with PKGame. The superiority of proposed strategy selection in terms of both complexity and efficiency is shown on real-world as well as artificial data sets. As a result, GTKMeans become suited for multi-objective clustering in large sized data sets as well as small and medium ones.

The future work includes studying the game-theoretic clustering approach in several directions: 1) considering mixed strategies in the approach as well as pure strategies, 2) using a precise method for adapting the strategy selection parameter.

**Acknowledgments**

The authors would like to acknowledge support from Iranians Research Institute for ICT. They would also like to thank extensive comments from the reviewers and the anonymous associate editor.

Table 1: Average execution time (seconds) of The Multi-Objective algorithms with and without using the strategy selection method on GTD for various cluster numbers.

| K | 4 | 5 | 6 | 7 | 8 |
|---|---|---|---|---|---|
| GTKMeans | 0.10 | 0.40 | 1.63 | 26.13 | 132.14 |
| SS(2) | **0.05** | **0.16** | **0.36** | **1.76** | **3.34** |
| SS(3) | **0.05** | **0.15** | **0.25** | **0.58** | **1.44** |
| PKGame | 0.04 | 0.07 | 0.45 | 7.25 | 10.14 |
| SS(2) | **0.01** | **0.04** | **0.10** | **0.47** | **0.65** |
| SS(3) | **0.02** | **0.03** | **0.06** | **0.15** | **0.22** |

Table 2: Average execution time (seconds) of PKGame with and without using the strategy selection method on DS1 for various cluster numbers.

| K | 4 | 5 | 6 | 7 | 8 |
|---|---|---|---|---|---|
| PKGame | 0.06 | 0.03 | 0.07 | 0.09 | 7.69 |
| SS(2) | **0.05** | **0.01** | **0.04** | **0.07** | **0.49** |
| SS(3) | **0.05** | **0.01** | **0.03** | **0.07** | **0.22** |

Table 3: Average execution time (seconds) of GTKMeans with and without using the strategy selection method on DS1 for various cluster numbers.

| K | 4 | 5 | 6 | 7 | 8 |
|---|---|---|---|---|---|
| GTKMeans | 0.35 | 6.28 | 24.05 | 109.86 | 118.42 |
| SS(2) | **0.21** | **0.55** | **1.31** | **3.27** | **6.15** |
| SS(3) | **0.19** | **0.23** | **0.47** | **1.07** | **1.11** |
| SS(4) | **0.19** | **0.16** | **0.31** | **0.70** | **0.68** |

Table 4: Jain's Fairness index and Geometric mean index for different size of clusters over DS1.

| | Jain's Fairness Index | | | | | Geometric Mean Index | | | | |
| K | 4 | 5 | 6 | 7 | 8 | 4 | 5 | 6 | 7 | 8 |
|---|---|---|---|---|---|---|---|---|---|---|
| GTKMeans | 0.9615 | 0.9907 | 0.9872 | 0.9518 | 0.9802 | 69.9695 | 65.1865 | 71.8699 | 70.9497 | 66.4784 |
| SS(2) | 0.9610 | 0.9900 | 0.9844 | 0.9508 | 0.9800 | 69.9490 | 65.0271 | 71.0536 | 70.6251 | 64.6521 |
| SS(3) | 0.9611 | 0.9900 | 0.9857 | 0.9509 | 0.9801 | 69.9593 | 65.0883 | 71.6349 | 70.6447 | 65.5762 |
| SS(4) | 0.9602 | 0.9901 | 0.9865 | 0.9501 | 0.9797 | 69.8914 | 65.0766 | 71.5910 | 70.3984 | 65.3661 |

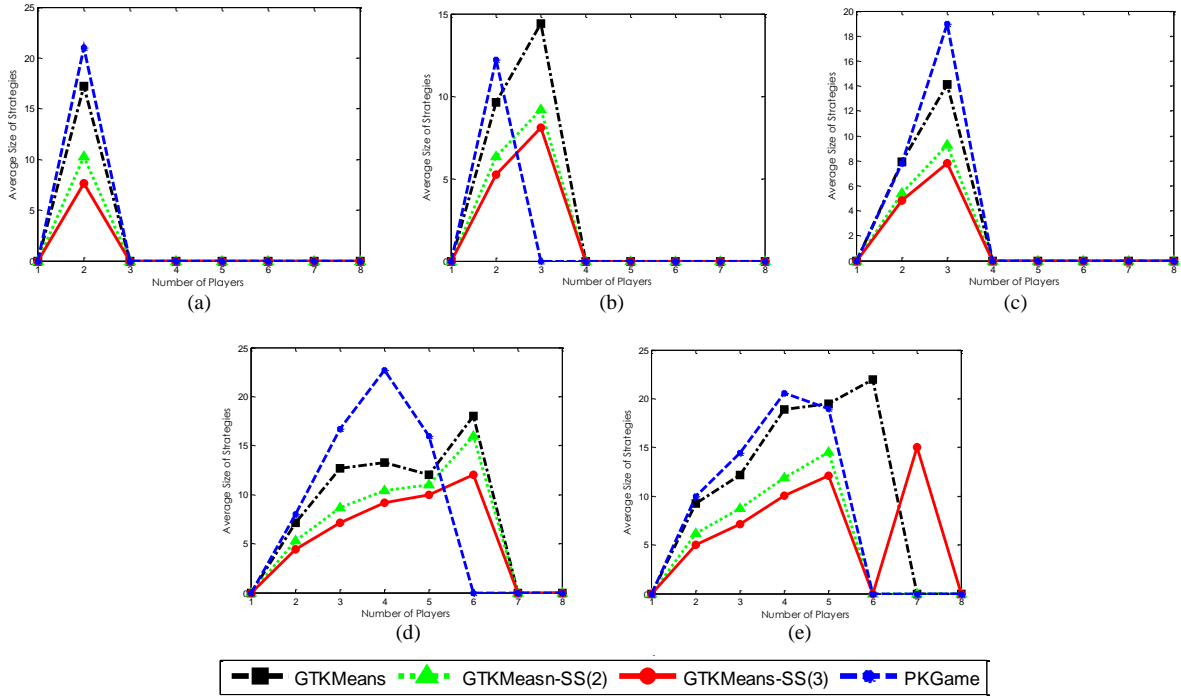

Figure 1. Analysis of GTKMeans, GTKMeans with strategy selection($N_s =2$), GTKMeans with strategy selection($N_s =3$), PKGame, PKGame with strategy selection($N_s =2$), PKGame with strategy selection($N_s =3$) on the basis of average number of strategies per players numbers. The experiments are performed on GTD for 50 iterations for different size of cluster number (K). (a) K=4 (b) K=5 (c) K=6 (d) K=7 (e) K=8

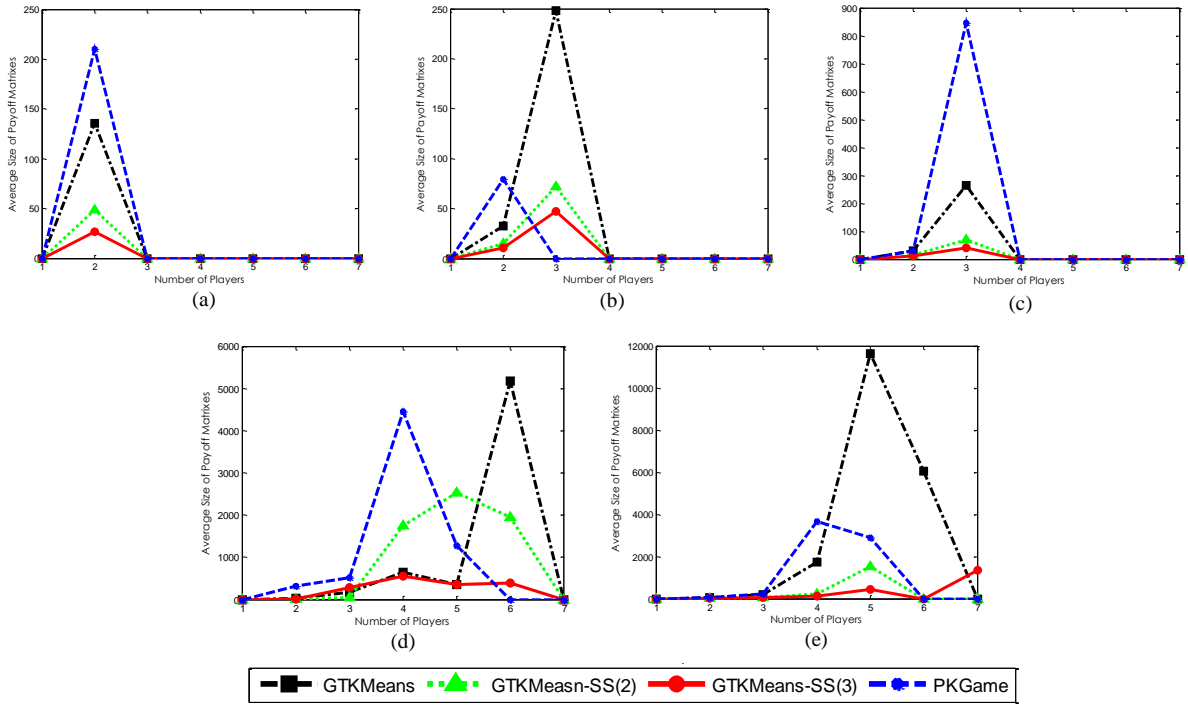

Figure 2. Analysis of GTKMeans, GTKMeans with strategy selection($N_s =2$), GTKMeans with strategy selection($N_s =3$), PKGame, PKGame with strategy selection($N_s =2$), PKGame with strategy selection($N_s =3$) on the basis of average size of payoff matrices per players numbers. The experiments are performed on GTD for 50 iterations for different size of cluster number (K). (a) K=4 (b) K=5 (c) K=6 (d) K=7 (e) K=8

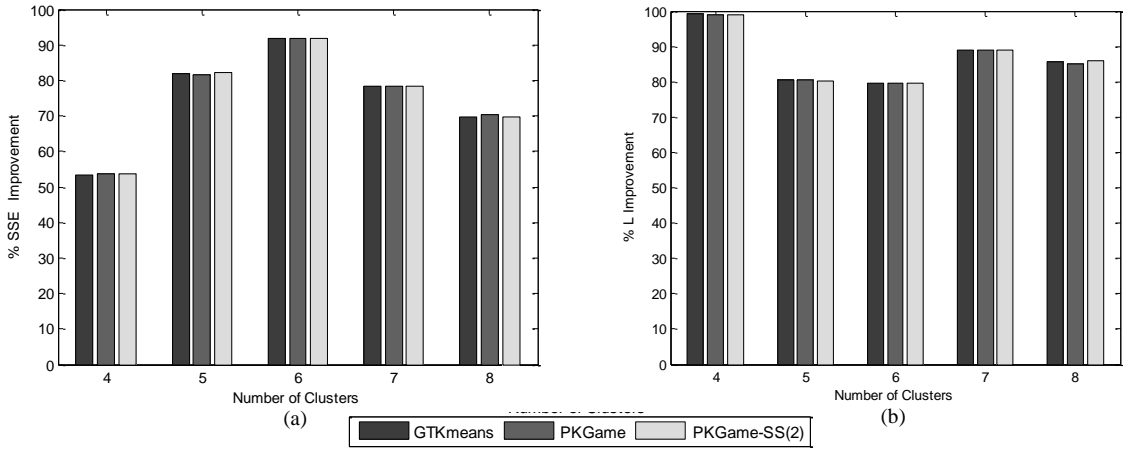

Figure 3. Performance comparison of GTKMeans and PKGame with and without using the strategy selection method over GTD.
(a) Improvement in SSE. (b) Improvement in L.

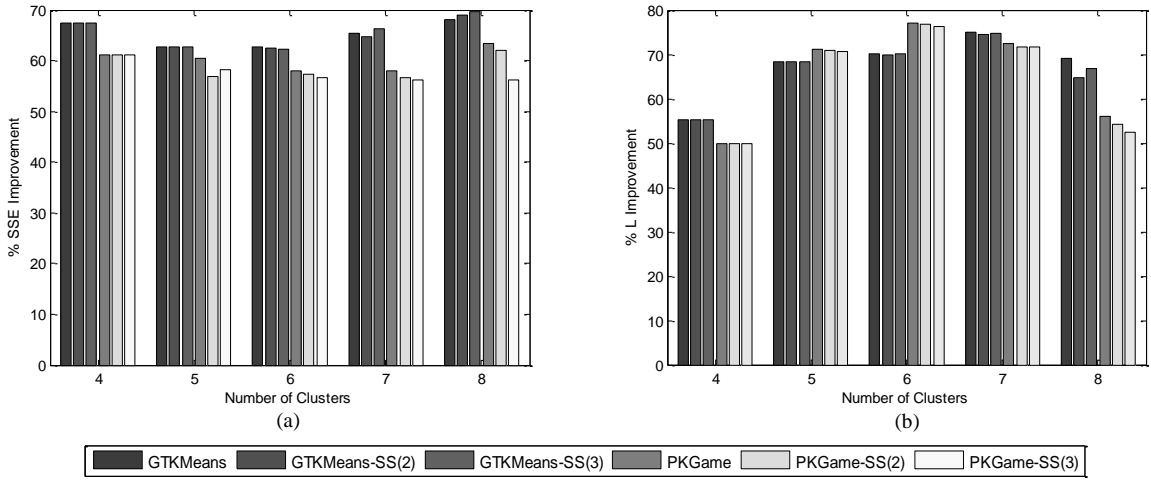

Figure 4. Performance comparison of PKGame with and without using the strategy selection method over BTD.
(a) Improvement in SSE. (b) Improvement in L.

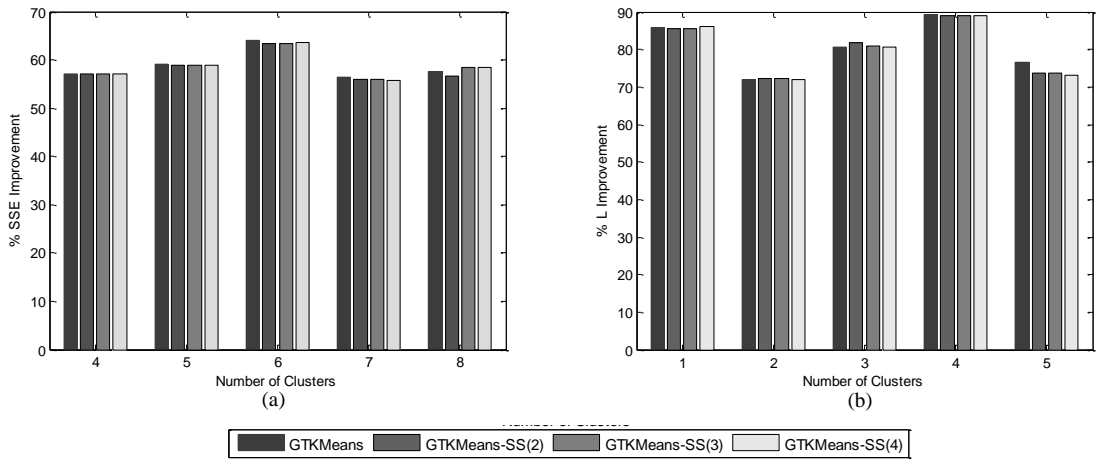

Figure 5. Performance comparison of GTKMeans with and without using the strategy selection method over DS1.
(a) Improvement in SSE. (b) Improvement in L.

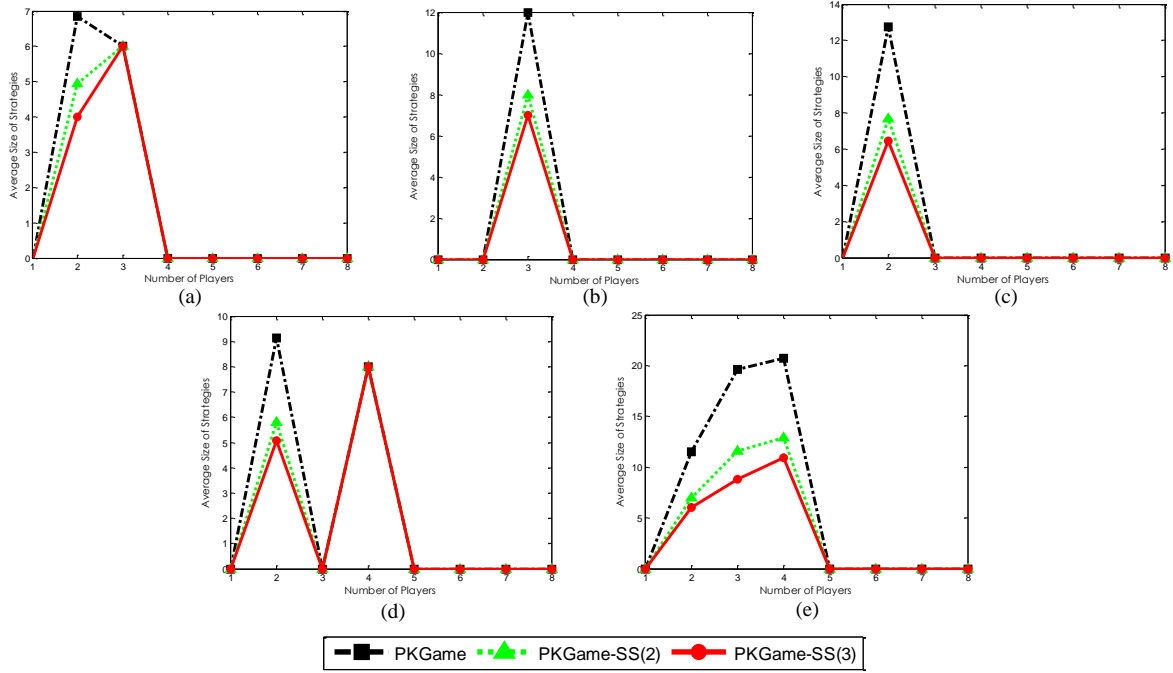

Figure 6. Analysis of PKGame, PKGame with strategy selection($N_s =2$), PKGame with strategy selection($N_s =3$), on the basis of average number of strategies per players numbers. The experiments are performed on DS1 for 50 iterations for different size of cluster number (K). (a) K=4 (b) K=5 (c) K=6 (d) K=7 (e) K=8

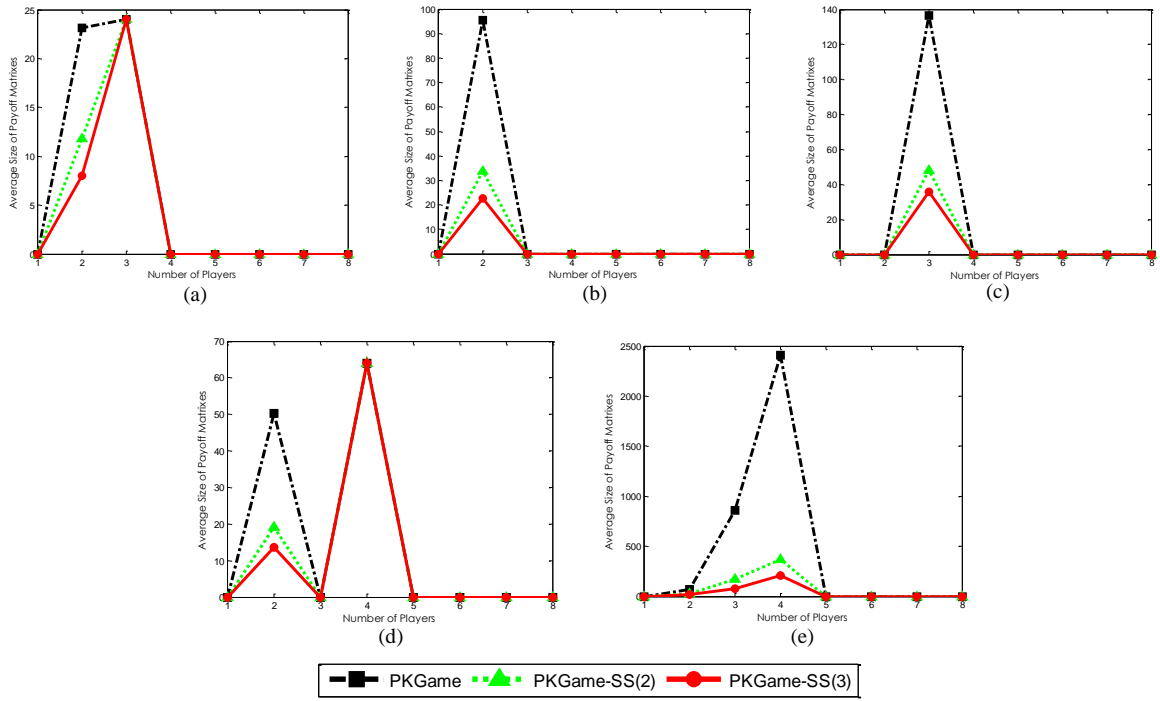

Figure 7. Analysis of PKGame, PKGame with strategy selection($N_s =2$), PKGame with strategy selection($N_s =3$), on the basis of average size of payoff matrices per players numbers. The experiments are performed on DS1 for 50 iterations for different size of cluster number (K). (a) K=4 (b) K=5 (c) K=6 (d) K=7 (e) K=8

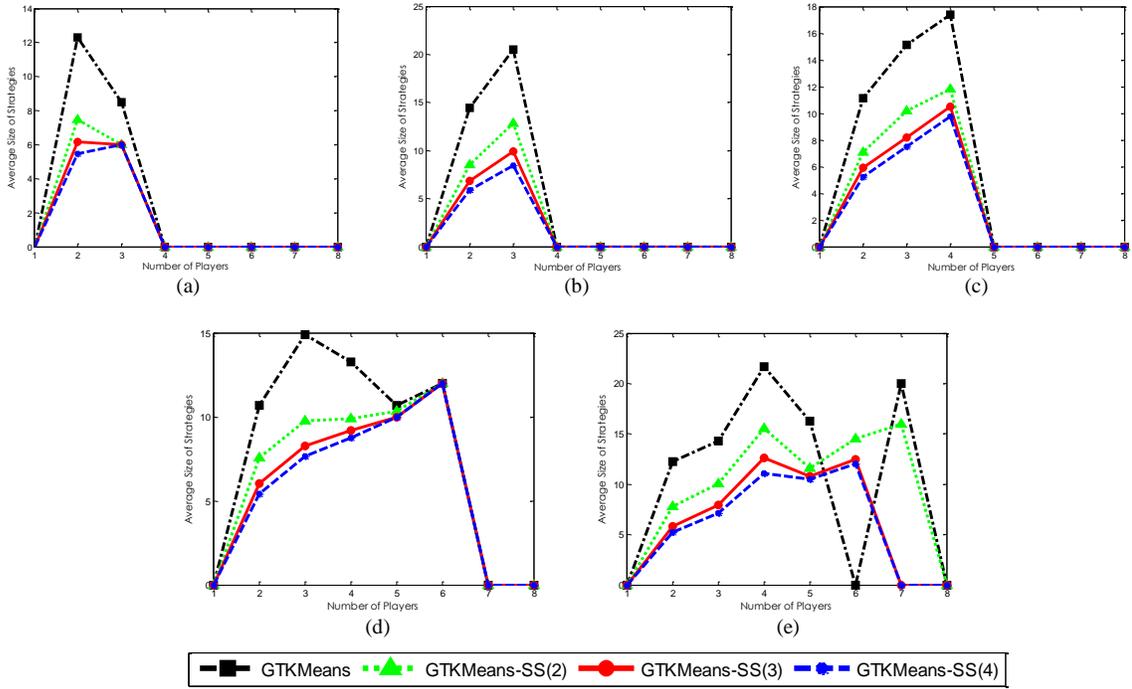

Figure 8. Analysis of GTKMeans, GTKMeans with strategy selection($N_s=2$), GTKMeans with strategy selection($N_s=3$) and GTKMeans with strategy selection($N_s=4$) on the basis of average number of strategies per players numbers. The experiments are performed on DS1 for 50 iterations for different size of cluster number (K). (a) K=4 (b) K=5 (c) K=6 (d) K=7 (e) K=8

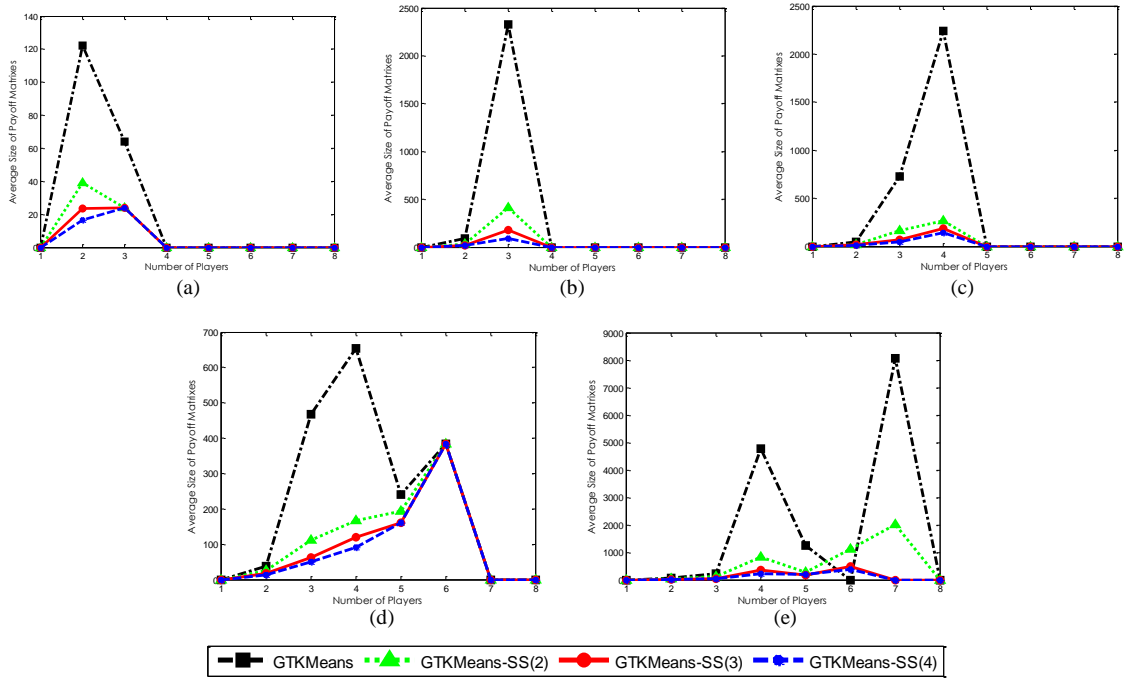

Figure 9. Analysis of GTKMeans, GTKMeans with strategy selection($N_s=2$), GTKMeans with strategy selection($N_s=3$) and GTKMeans with strategy selection($N_s=4$), the basis of average size of payoff matrices per players numbers. The experiments are performed on DS1 for 50 iterations for different size of cluster number (K). (a) K=4 (b) K=5 (c) K=6 (d) K=7 (e) K=8